# Artificial Intelligence as an Economic, Environmental, Geopolitical, and Social Transformation

**Marcin Marciniak**

**Institute of Theoretical Physics and Astrophysics**

**University of Gdańsk**

*Email: marcin.marciniak@ug.edu.pl*

**Abstract**

The rapid development of artificial intelligence is often discussed primarily as a technological breakthrough. Such an approach is insufficient because AI is also transforming the allocation of capital, energy, natural resources, labour, and political power. Investment in generative AI and computing infrastructure is increasing rapidly and is increasingly concentrated among a small number of corporations and countries. At the same time, the expansion of data centres creates new electricity and water demands, potentially producing local infrastructure bottlenecks and distributive conflicts. The AI value chain is dependent on geographically concentrated supplies of advanced semiconductors, manufacturing equipment, cloud services, energy, and specialised knowledge. Control over these bottlenecks may give states and corporations structural power that does not derive directly from military superiority. The environmental effects of AI are similarly ambivalent: AI systems consume energy, water, materials, and computing hardware, but may also improve energy efficiency, climate modelling, renewable-energy integration, and environmental monitoring. Finally, AI-enabled automation, including intelligent robotics, may reduce employment in some occupations while augmenting labour and creating new tasks in others. The social outcome of this transformation will therefore depend not only on the technology itself but also on competition policy, infrastructure planning, environmental regulation, social protection, collective bargaining, and the distribution of productivity gains.

**Keywords:** artificial intelligence, automation, robotics, energy demand, semiconductors, employment, environment, geopolitics, trade unions

## 1. Introduction

Artificial intelligence is becoming a general-purpose technology whose effects extend far beyond the information technology sector. Recent advances in machine learning, generative AI, computer vision, and robotics have accelerated investment in computing infrastructure and broadened the range of tasks that can be performed or supported by machines. AI is already being applied in manufacturing, logistics, finance, healthcare, scientific research, public administration, energy systems, and creative industries.

The dominant public debate tends to alternate between technological optimism and predictions of large-scale social disruption. The optimistic position emphasises higher productivity, scientific progress, safer working conditions, and more efficient use of resources. The pessimistic position focuses on job displacement, corporate concentration, environmental costs, surveillance, misinformation, and geopolitical conflict. Neither perspective adequately captures the conditional character of AI's effects.

The consequences of AI are not determined by technical capabilities alone. They also depend on the ownership of infrastructure, access to energy and computing power, the structure of product and labour markets, public regulation, and the ability of social institutions to distribute the costs and benefits of technological change.

This article advances the following argument:

The development of artificial intelligence constitutes an accelerated reallocation of capital, energy, natural resources, labour, and political power. Its economic, environmental, geopolitical, and social balance is not predetermined but will depend on the institutions governing infrastructure, markets, employment, and international interdependence.

The article examines five interconnected dimensions of this transformation: investment and market concentration, energy demand, geopolitical bottlenecks, environmental consequences, and employment - particularly the effects of AI-enabled robotics on traditional industrial occupations.

## 2. Investment, Productivity, and Market Concentration

Private investment in generative AI and related computing infrastructure has grown rapidly. According to the Stanford Institute for Human-Centered Artificial Intelligence, private investment in generative AI reached USD 33.9 billion in 2024, an increase of 18.7 per cent over the previous year and more than eight times the 2022 level (Stanford HAI, 2025).

However, the proposition that investment in AI is growing faster than investment in every other sector should be treated cautiously. Venture-capital investment, corporate capital expenditure, public research funding, and total fixed-capital formation are different statistical categories. Comparisons based on incompatible measures may exaggerate the exceptional character of AI investment.

A more defensible conclusion is that expenditure on generative AI, data centres, advanced semiconductors, and associated energy infrastructure constitutes one of the fastest-growing categories of technological investment. This expansion has four important economic consequences.

First, high development costs favour the concentration of AI capabilities. Training a frontier model requires advanced processors, large datasets, specialised labour, electricity, and access to cloud infrastructure. The cost is incurred before the commercial value of the model is known, while repeated training, evaluation, deployment, and model updating create continuing expenditure. UNCTAD therefore treats infrastructure, data, and skills as strategic leverage points in the emerging AI economy (UNCTAD, 2025). Stanford HAI reports that nearly 90 per cent of notable AI models released in 2024 came from industry, compared with 60 per cent in 2023 (Stanford HAI, 2025).

Concentration is reinforced by vertical links between semiconductors, cloud platforms, foundation models, and the products through which models reach users. A company that already owns a global cloud platform, a large customer base, and a profitable digital business can reserve scarce processors, spread fixed costs across many services, provide cloud credits to developers, and integrate its own model into established software. New firms may innovate at the application layer, but they often remain dependent on infrastructure supplied by a small number of incumbents. European competition authorities have consequently identified access to compute, cloud services, data, and distribution channels, as well as interoperability and switching costs, as potential sources of foreclosure and lock-in (European Commission, 2024).

The resulting increase in the power of Big Tech is economically and politically dangerous even if it does not immediately produce a conventional monopoly. Control over essential infrastructure can influence which firms are able to enter the market, which research questions can be pursued, which technical standards become dominant, and on what terms governments, universities, and smaller businesses obtain access to AI. It can also allow the same companies to act simultaneously as infrastructure providers, model developers, investors, and competitors of their own customers. This creates conflicts of interest and makes effective competition more difficult to assess.

The conclusion should nevertheless be qualified. Falling inference costs, smaller models, open-weight releases, and rented cloud capacity can reduce some entry barriers. In 2024, the performance gap between leading closed and open-weight models narrowed substantially on several benchmarks (Stanford HAI, 2025). These developments make the market more contestable, but they do not eliminate dependence on advanced chips, hyperscale clouds, distribution platforms, and the financial capacity to train and operate systems at scale. Competition policy must therefore examine the entire supply chain rather than only the number of visible model providers.

Second, AI investment has an opportunity cost. Capital, engineering capacity, electricity infrastructure, water, land, and public subsidies directed towards AI are not simultaneously available for all other uses. Opportunity cost does not mean that every AI project displaces a demonstrably superior project. It means that investment decisions must be compared with realistic alternatives and assessed in terms of their full social return rather than their private return alone.

Ireland illustrates the issue at the level of national infrastructure planning. Data centres accounted for 22 per cent of Irish electricity demand in 2024, and contracted demand could raise their share to 31 per cent by 2034. The Commission for Regulation of Utilities responded by requiring connection decisions to take account of network constraints and by obliging new data centres to provide matching generation or storage capacity and additional renewable supply (CRU, 2025). The relevant opportunity cost is not captured by claiming that a particular unit of electricity would otherwise have gone to a hospital or a household. It arises because grid capacity, engineering effort, generation, and finance committed to one rapidly growing class of load cannot be deployed at the same time for housing, manufacturing, transport electrification, or heating. Chapter 3 examines these infrastructure trade-offs in greater detail.

Chile provides a parallel example involving water rather than electricity. In February 2024, the Second Environmental Court partially annulled the approval of Google's proposed Cerrillos data centre and required the environmental assessment to incorporate the effects of climate change on the Central Santiago Aquifer. Google subsequently withdrew the original design and announced that a new proposal would use air cooling (Second Environmental Court of Chile, 2024). Research on the Cerrillos controversy describes a conflict between a national narrative of digital modernisation and local concerns about drought, water security, and the distribution of environmental costs (Tironi and Albornoz, 2025). The case demonstrates that the relevant choice is often not simply between building and not building. Alternative locations, cooling technologies, water sources, and compensation arrangements can change both the private cost of the project and its social opportunity cost.

Universities face a third form of opportunity cost: the allocation of research budgets and the direction of knowledge production. Ahmed and Wahed analysed 171,394 papers from 57 major computer-science conferences and found that the rise of deep learning increased the representation of large technology firms and elite universities while mid-ranked and lower-ranked universities lost ground. They identify unequal access to computing resources as an important mechanism (Ahmed and Wahed, 2022). Besiroglu et al. similarly document a declining representation of academic-only teams in compute-intensive fields, particularly foundation-model research, and warn that the divide may reduce independent scrutiny of influential systems (Besiroglu et al., 2024). For a university, spending on GPU clusters or commercial cloud services competes with other laboratories, disciplines, staff, and teaching needs. Dependence on industry-provided models or cloud credits can also shift the research agenda towards questions that fit corporate infrastructure and away from independent replication, safety testing, or less commercially attractive problems.

These examples do not establish that AI infrastructure has a lower social return than the alternatives. Data centres can support digital services, universities require modern computing capacity, and more efficient cooling can reconcile investment with environmental limits. They do show why public support should be accompanied by transparent cost-benefit analysis, realistic forecasts, environmental assessment, access conditions, and rules specifying who pays when a project requires complementary infrastructure.

Third, rapid investment growth may create financial risk. Expectations concerning future AI revenues may exceed the rate at which firms can develop commercially valuable applications and customers can reorganise production around them. High valuations and capital expenditure can be rational if AI becomes a widely used general-purpose technology, yet the same expectations can produce duplicated facilities, overestimated demand, and investments whose private returns are too low to justify their cost.

The telecommunications and internet boom of the late 1990s offers a useful historical comparison. Regulatory change, rapidly improving fibre-optic technology, abundant finance, and optimistic forecasts of internet traffic encouraged competing carriers to build long-distance networks. In the United States, capital expenditure by publicly traded telecommunications service companies rose from USD 47 billion in 1995 to USD 121 billion in 2000, before falling to USD 49 billion in 2002. WorldCom, Global Crossing, and many smaller companies failed, and the contraction in telecommunications investment accounted for a large part of the decline in US information-technology investment after 2000 (Doms, 2004).

What the comparison shows is that a transformative technology and an investment bubble are not mutually exclusive. Demand for digital communication continued to grow after the crash, while fibre networks built during the boom became part of the infrastructure of the later internet economy. Research by Hogendorn complicates the familiar story of wholly wasteful construction: once swaps and leases between networks are separated from physically owned route miles, excessive entry appears less extensive than headline measures suggested (Hogendorn, 2011). Perez places this pattern in a broader account of technological revolutions, in which speculative finance can accelerate an initial infrastructure build-out before a correction separates viable applications from failed business models (Perez, 2002).

The analogy also has limits. Fibre is a relatively durable and general network asset. AI accelerators depreciate quickly, are replaced by more efficient generations, and may be specialised for particular software ecosystems. A data centre can remain useful after an AI correction, but the processors inside it, the power contracts supporting it, or the location chosen for it may not retain their expected value. Conversely, much of the present investment is undertaken by profitable technology companies with large cash flows rather than by highly leveraged telecommunications entrants, and current AI systems already generate observable demand and productivity gains. The historical evidence therefore supports neither the claim that a crash is inevitable nor the claim that all infrastructure will remain socially valuable after one.

The appropriate lesson is institutional. Forecasts should be stress-tested, investment should be staged where possible, and the parties creating project-specific risk should bear a corresponding share of the cost. Regulators should distinguish between reusable infrastructure and assets whose value depends on one developer, one model architecture, or one exceptionally optimistic demand forecast. Otherwise, private investors may retain the upside while utilities, public authorities, workers, or local communities absorb stranded costs after a correction.

Fourth, productivity gains are likely to be distributed unevenly. AI may reduce production costs and improve worker performance, but increased productivity does not automatically translate into higher wages, shorter working hours, or greater employment security. Distribution depends on whether AI complements or substitutes particular tasks, who owns the system and the data used to improve it, the degree of product-market competition, and the bargaining power of workers.

Controlled studies provide credible evidence of substantial task-level gains. In an experiment involving 453 college-educated professionals performing writing tasks, access to ChatGPT reduced completion time by 40 per cent and increased assessed output quality by 18 per cent. The productivity distribution also narrowed because participants who initially performed less well benefited more (Noy and Zhang, 2023). This is evidence that generative AI can improve performance and reduce skill gaps within a defined set of tasks. It is not evidence that the resulting economic surplus will be paid to workers.

Evidence from an actual workplace leads to a similar but more nuanced conclusion. Brynjolfsson, Li, and Raymond studied 5,172 customer-support agents and found that AI assistance increased issues resolved per hour by 15 per cent on average. Less experienced and lower-skilled workers improved most, whereas highly skilled workers obtained small gains in speed and, in some cases, a slight decline in quality. The system also reduced worker attrition and improved customer interactions. The authors explicitly caution, however, that a study within one firm does not determine aggregate wage or employment effects (Brynjolfsson, Li, and Raymond, 2025). A company may share the gain through wages or reduced working time, use it to expand output and employment, or use it to reduce staffing while retaining the surplus.

Other markets already show potential losses. Hui, Reshef, and Zhou found that freelancers in occupations highly exposed to text- and image-generating systems experienced reductions in both employment and earnings after the release of ChatGPT, DALL-E 2, and Midjourney. Strong past performance did not protect workers, and the evidence suggested that some top freelancers were disproportionately affected (Hui, Reshef, and Zhou, 2024). This does not establish the long-run effect of AI on all creative work, but it shows how productivity-enhancing technology for clients can simultaneously reduce demand and income for suppliers of substitutable tasks.

The distribution of gains also operates through firms and ownership. Babina et al. find that US firms investing more intensively in AI experienced faster growth in sales, employment, market valuation, and product innovation, but that the gains were concentrated among larger firms and were associated with greater industry concentration (Babina et al., 2024). At the macroeconomic level, models of automation show that technical change can increase returns to wealth while wages at the lower end of the distribution stagnate, thereby widening income and wealth inequality even when aggregate productivity rises (Moll, Rachel, and Restrepo, 2022).

The policy question is therefore not whether AI raises productivity in the abstract, but how the surplus is governed. Competition policy can limit rents arising from market power; taxation can redistribute part of the return to capital; collective bargaining and worker participation can influence wages, staffing, surveillance, and working time; training can help workers move into complementary tasks; and social protection can reduce the cost of displacement. Without such institutions, the same technological improvement can produce higher output together with weaker job security, greater concentration, and a declining labour share. With them, productivity growth can support higher wages, better services, safer work, and shorter hours.

**3. Energy Demand and Infrastructure Constraints**

There is no AI at scale without electricity.

Training and operating AI systems require data centres containing processors, memory, networking equipment, cooling systems, and backup power. The International Energy Agency estimates that data centres consumed approximately 415 terawatt-hours of electricity in 2024, equivalent to about 1.5 per cent of global electricity consumption, and expects their demand to more than double by 2030 in its base-case scenario (IEA, 2025). Not all data-centre consumption is attributable to AI: these facilities also provide cloud computing, storage, streaming, financial, and other digital services. Nevertheless, AI is becoming an important driver of new capacity because AI-oriented facilities tend to contain unusually dense and power-intensive computing equipment.

These projections do not establish that AI will cause a global shortage of electricity. The more immediate risk is a mismatch between the location and speed of new demand and the availability of generation, substations, transmission lines, and distribution networks. Electricity may be available at the national level while a particular region lacks the infrastructure required to deliver it. Large power plants can also sometimes be planned or contracted more quickly than high-voltage lines, which require permits, land, equipment, and lengthy construction. The relevant constraint is therefore not only the total amount of electricity produced, but whether sufficient capacity is available at the required place and time.

The effect can already be observed in housing and business development. In West London, particularly in Ealing, Hillingdon, and Hounslow, the expansion of data centres along the M4 corridor contributed to severe grid constraints. A London Assembly report records that, from 2022, some housing developments were warned that they might have to wait until 2037 for an electricity connection. The first-come, first-served allocation of grid capacity delayed both housing construction and business growth (Greater London Authority, 2025). This does not mean that data centres were the only cause of the shortage; population growth, electrification, and other development also increased demand. It does show how a concentrated cluster of large consumers can absorb scarce local connection capacity and postpone projects with very different social purposes.

Ireland provides a closely related example of transmission congestion and security-of-supply concerns. Electricity demand from data centres increased from 5 per cent of national demand in 2015 to 22 per cent in 2024. On the basis of already contracted demand, the Commission for Regulation of Utilities projects that the share could reach 31 per cent by 2034 (CRU, 2025). Ireland's connection policy now requires system operators to consider whether a proposed facility is located in a constrained part of the network. New data centres must also provide generation or storage capacity corresponding to their maximum requested import capacity and develop additional renewable supply. These measures illustrate that the problem is not simply annual energy consumption. A large facility must also be supported during periods of peak demand and system stress.

Grid congestion creates infrastructure costs that extend beyond the connecting customer. New substations, transmission corridors, generation capacity, and reserve resources must often be built before the final level of demand is known. Virginia illustrates the scale of this problem. An independent forecast commissioned by the state's Joint Legislative Audit and Review Commission found that unconstrained electricity demand could double within ten years, with data centres as the principal driver. Meeting even half of this demand would require substantial additions of generation and transmission capacity and would be difficult under both fossil-intensive and cleaner-generation scenarios (JLARC, 2024).

The same Virginia study also shows why infrastructure expansion can place pressure on electricity tariffs. It found that data centres were paying their full cost of service under the existing rate structure; the report therefore did not identify a simple current cross-subsidy. It nevertheless estimated that the wider generation and transmission investment required by future demand, together with higher energy prices and greater exposure to imported electricity, could increase the generation- and transmission-related costs of a typical Dominion Energy residential customer by approximately USD 14-37 per month in real terms by 2040 (JLARC, 2024). This is a scenario-based projection rather than an observed increase, but it demonstrates how system-wide costs can reach households even when a large consumer pays the charges directly assigned to it.

Limited grid capacity also creates competition between alternative uses of electricity. Data centres are not competing only with existing household consumption. They may seek connections at the same time as manufacturers, public services, housing developments, electric-vehicle charging infrastructure, and heat pumps. Denmark made this conflict explicit in 2026 when the government proposed replacing the first-come, first-served connection rule with a system based on societal priority. Under the proposal, healthcare, defence, businesses, and households seeking connections for electric-vehicle chargers or heat pumps could be placed ahead of large data centres (Eurofound, 2026). Because the measure was proposed as a preventive response, it should not be treated as proof that data centres had already displaced all these uses. It is, however, clear evidence that governments may have to make distributive choices about access to scarce network capacity.

Rapid demand growth can also prolong the use of fossil-fuel generation. Georgia's approved 2025 Integrated Resource Plan combined renewable energy, battery storage, transmission investment, and demand-side measures with the continued operation of coal capacity and additional gas resources. The Bowen coal plant, previously expected to close in 2035, may continue operating until 2038 as part of the response to anticipated demand, much of which is associated with data-centre development (Georgia Power, 2025; Walton, 2025). The example should be interpreted carefully: data centres are not the sole cause of Georgia's generation decisions, and the plan contains substantial non-fossil investment. It nevertheless demonstrates a real mechanism through which unexpectedly rapid load growth can delay plant retirement or justify new fossil capacity when cleaner resources and networks cannot be delivered at the same speed.

A further conflict concerns who should bear the financial risk of grid expansion. Utilities may have to invest before a data-centre project is completed. If the project is cancelled, delayed, or uses much less electricity than announced, the network may be left with stranded assets whose costs would otherwise be recovered from existing customers. Ohio addressed this problem through a dedicated AEP Ohio data-centre tariff, approved by the Public Utilities Commission of Ohio in July 2025. The tariff applies long-term minimum demand charges, collateral requirements, and exit fees to very large new loads. Its contract period includes a load-ramp phase of up to four years followed by eight years, while monthly charges are linked to a substantial proportion of contracted or previously reached demand (AEP Ohio, 2025; PUCO, 2025). The explicit objective is to prevent residential, commercial, and industrial customers from bearing the cost of underused infrastructure constructed for projects that fail to materialise.

These cases reveal a broader planning problem. Data-centre load forecasts can be unusually uncertain because several developers may request connections for competing sites, while the same expected AI demand may appear in the plans of several regions. Building too little infrastructure can delay investment and threaten reliability; building too much can create stranded assets and long-term costs for consumers. Regulators therefore need evidence that projects are financially credible, realistic connection schedules, staged capacity commitments, and rules specifying who pays if forecast demand does not appear.

The relationship between AI and energy is nevertheless bidirectional. AI can improve demand forecasting, predictive maintenance, grid balancing, energy trading, and the integration of variable renewable generation. The IEA estimates that widespread adoption could unlock additional effective transmission capacity and produce energy savings in some industrial sectors (IEA, 2025). These estimates describe scenario-dependent opportunities, not guaranteed outcomes. They depend on data quality, system integration, incentives, cybersecurity, and the willingness of operators to act on model recommendations.

Efficiency improvements may also be offset by a rebound effect. If better processors and models reduce the electricity required for an individual AI operation, lower costs may lead firms to use AI much more frequently and in a wider range of products. Declining energy consumption per query can therefore coexist with rising total consumption. The appropriate policy question is not simply whether AI consumes electricity, but whether the social value of particular applications justifies their marginal use of energy and infrastructure, whether cleaner alternatives can be deployed in time, and whether the associated costs and risks are distributed fairly.

## 4. The Environmental Balance of AI

The environmental effects of AI cannot be evaluated solely through the electricity consumed during model training. A comprehensive assessment must include the entire lifecycle of the system:

1. extraction and processing of raw materials;
2. semiconductor and server manufacturing;
3. construction of data centres;
4. model development, experimentation, and training;
5. inference, storage, and daily operation;
6. cooling and water consumption;
7. replacement, reuse, and transport of computing equipment;
8. electronic-waste treatment and recovery of materials.

The United Nations Environment Programme has therefore called for end-to-end environmental assessment of AI systems, while the OECD recommends lifecycle-based measurement covering electricity, greenhouse-gas emissions, water consumption, and embodied impacts (UNEP, 2024; OECD, 2022). The point is not merely to compile a longer list of burdens. It is to prevent an improvement at one stage from being reported as an overall environmental gain when costs have been shifted to another stage, place, or population.

A useful starting framework distinguishes three channels. The first is the direct footprint of computing infrastructure. The second consists of the effects enabled by AI applications in energy, transport, buildings, industry, agriculture, and environmental monitoring. The third comprises system-level changes in prices, behaviour, production, and institutions. Kaack et al. use a closely related distinction between computing-related impacts, immediate application impacts, and system-level impacts on greenhouse-gas emissions (Kaack et al., 2022). The environmental balance is the combined result of all three channels, not a subtraction of unrelated global estimates.

In practical terms, a positive application should be compared with a realistic counterfactual: the same activity, over the same period, without the AI system. The relevant net effect equals the resources and emissions genuinely avoided because of the application, minus the operational and lifecycle footprint of the additional digital system, and minus any rebound or leakage it induces. This formulation makes the causal question explicit. An AI system for detecting methane leaks produces a benefit only if it identifies leaks that would otherwise have persisted and if repairs follow. A traffic-optimisation system has a positive balance only if its energy savings are not offset by additional journeys or a shift away from public transport.

Several mechanisms can improve this balance.

First, measurement and disclosure make environmental effects visible at the level at which decisions are taken. Developers should report energy use, electricity source and location, carbon emissions, water withdrawal and consumption, hardware type, utilisation, and the number of repeated training or evaluation runs. Results should also be expressed per useful output—such as a completed inference task at a stated quality level—so that systems of different sizes can be compared. Location- and time-specific data are important because identical computations can have different emissions and water impacts depending on grid conditions, climate, and local scarcity.

The European Union has begun to institutionalise this information layer. Commission Delegated Regulation (EU) 2024/1364 requires operators of data centres with installed information-technology power demand of at least 500 kW to report harmonised indicators concerning energy consumption, renewable energy, waste-heat reuse, cooling, and water use (European Union, 2024). Reporting is necessary but not sufficient: it changes outcomes only when the information affects procurement, planning permission, grid-connection terms, financing, or minimum-performance standards.

Second, computational demand can be reduced before energy supply is considered. The appropriate model is not always the largest available model. Reusing an existing model, selecting a smaller specialised model, limiting hyperparameter searches, debugging at small scale, improving data quality, batching requests, caching repeated outputs, and applying distillation, pruning, or quantisation can reduce the computation needed for a given service. The Green AI approach therefore proposes treating efficiency as an evaluation criterion alongside accuracy rather than rewarding performance improvements without regard to computational cost (Schwartz et al., 2020).

The relationship between speed and environmental efficiency is not automatic. Lannelongue, Grealey, and Inouye show that, in one parallel-computing example, increasing the number of cores from 15 to 60 halved the running time but approximately doubled emissions because the extra hardware was used inefficiently. Their Green Algorithms framework combines runtime, processors, memory, facility overhead, utilisation, and location to estimate the operational carbon footprint of a computation (Lannelongue et al., 2021). This illustrates why procurement and research assessment should reward useful performance per unit of resource, not scale or speed alone.

Efficiency, however, balances the environmental impact only if absolute demand is controlled. Lower cost per inference can encourage more users, longer prompts, automated generation at scale, and the integration of AI into activities that previously required no comparable computation. Total electricity or water use may therefore rise even when every individual request becomes more efficient. This rebound mechanism is the reason to combine unit-efficiency targets with absolute energy, emissions, or water budgets where local or climate constraints require them.

Third, flexible computation can be matched more closely to low-carbon electricity. Some training, batch inference, data processing, and model evaluation can be shifted to hours or locations with lower grid carbon intensity. Radovanović et al. describe an operational system that forecasts next-day carbon intensity and delays flexible workloads during carbon-intensive hours while preserving their daily completion requirement (Radovanović et al., 2023). Similar scheduling can also reduce peak demand and the need for seldom-used network or generation capacity.

Temporal and spatial shifting cannot replace additional clean supply. It is most effective when workloads are genuinely flexible, transmission is available, and the shift changes which generators operate at the margin. Annual renewable-energy certificates may improve contractual accounting without ensuring that clean electricity is available in the hours and locations in which a data centre consumes power. A stronger balancing mechanism combines flexible demand with additional low-carbon generation, storage, network reinforcement, and rules requiring the beneficiary of a large new load to bear an appropriate share of those costs.

Fourth, energy and water must be optimised jointly. Evaporative cooling can reduce electricity used for cooling but consume substantial freshwater; dry cooling can greatly reduce on-site water use while increasing electricity demand. Carbon-aware scheduling may favour sunny hours, whereas water-aware scheduling may favour cooler hours. Li et al. show that carbon and water intensities vary differently across time and place and warn that minimising one footprint can worsen the other (Li et al., 2025). The appropriate objective is therefore not the smallest number of litres or kilowatt-hours in isolation, but the lowest combined impact under local water stress, grid conditions, and climate.

Practical measures include locating water-intensive facilities away from stressed catchments, using reclaimed rather than potable water where safe, adopting closed-loop or low-water cooling, adjusting operations during drought and heat, and publishing both withdrawal and consumption. The distinction matters: withdrawal measures competition for a shared source, while consumption measures water that is not promptly returned to the same catchment. Local limits may justifiably override a globally favourable carbon calculation when a project threatens municipal, agricultural, or ecological water security.

Fifth, the embodied footprint of hardware can be balanced through longer service life and circular use. More efficient processors reduce operational energy, but frequent replacement increases semiconductor production, transport, raw-material extraction, and electronic waste. The optimal replacement date therefore depends on whether the operating savings of new hardware exceed its additional embodied impacts. Modular repair, reuse of older accelerators for less demanding workloads, refurbishment, supplier take-back, material recovery, and procurement requirements for durability can reduce this burden. Lifecycle accounting is essential because operational metrics such as power-usage effectiveness do not capture it (UNEP, 2024; OECD, 2022).

Sixth, beneficial applications require measurement of realised rather than theoretical savings. AI can improve renewable-generation forecasts, optimise electricity networks, control heating and cooling, detect methane leaks, reduce industrial energy use, and support materials discovery. The IEA estimates that widespread adoption of already identified AI applications could reduce emissions by 1,400 million tonnes of $CO_2$ in 2035—several times the projected emissions of data centres in its scenarios. This is an exploratory widespread-adoption case, not a forecast. The IEA emphasises that data, infrastructure, skills, regulation, and implementation barriers may prevent the savings from materialising (IEA, 2025).

The same application can also have opposite system-level effects. More efficient oil and gas exploration can reduce energy used per unit while extending fossil-fuel production. Autonomous vehicles can improve routing while drawing passengers away from public transport. Cheaper industrial production can expand total output enough to offset efficiency gains. The IEA explicitly notes that rebound effects may negate part of the potential emissions reductions, and Kaack et al. similarly stress that market and policy conditions determine whether AI-enabled optimisation supports or obstructs mitigation (IEA, 2025; Kaack et al., 2022). Assessment should therefore monitor total sectoral outcomes after deployment, not stop at a laboratory benchmark or an engineering estimate.

Seventh, economic and regulatory rules determine who has an incentive to act on these mechanisms. Carbon and water prices can internalise part of the environmental cost; performance standards can remove the least efficient facilities; public procurement and subsidies can require lifecycle reporting and verified net benefits; connection agreements can require demand flexibility, additional generation or storage, and contributions to grid expansion; and environmental permits can impose catchment-specific water limits. Community participation and compensation are also relevant where local residents bear land, noise, water, or infrastructure costs while benefits accrue elsewhere.

A defensible decision hierarchy follows from these mechanisms. Avoid unnecessary computation first; use the least resource-intensive system capable of delivering the required result; improve hardware and software efficiency; schedule flexible workloads in environmentally favourable hours and places; supply the remaining demand with additional low-carbon energy and appropriate cooling; extend and circularise the hardware lifecycle; and offset residual impacts only after direct reductions have been exhausted. Environmental claims should then be verified against absolute outcomes and a stated counterfactual.

The positive and negative effects of AI are therefore not self-balancing. A beneficial application in one sector does not grant a general environmental credit to all AI infrastructure, and lower resource use per task does not guarantee a fall in total consumption. Balance is produced by measurement, technological choices, demand management, lifecycle design, and institutions that link private decisions to public costs. The direct footprint of AI, the effects of particular applications, and wider systemic consequences should remain analytically separate and then be combined only within a transparent, location-specific, and counterfactual assessment.

## 5. Semiconductor Bottlenecks and Geopolitical Power

The physical infrastructure of AI is not produced by a single, linear chain. It is assembled through a network of specialised industries that links chip architecture and design software to manufacturing equipment, materials, wafer fabrication, memory, advanced packaging, server assembly, high-speed networking, and cloud operation. The OECD notes that semiconductor production can involve more than one thousand process steps and, for some integrated circuits, as many as 500 specialised chemicals. A disruption in one input can therefore halt output far downstream even when all other components are available (OECD, 2023b; OECD, 2025).

The word 'largest' must be used carefully in this context. Companies operating at different stages cannot be ranked meaningfully by one measure of size: revenue, fabrication capacity, technological capability, and control over a non-substitutable input are different forms of importance. The following firms should therefore be understood as the leading or strategically most significant actors within particular segments, rather than as one universal league table.

A useful way to understand the chain is to follow an AI accelerator from its specification to its use in a data centre. The process begins with architecture, reusable intellectual property, and electronic design automation; proceeds through the design of the accelerator and its surrounding system; moves to the production of silicon wafers and memory; and ends with packaging, testing, server integration, networking, and deployment by a cloud or data-centre operator. The firms coordinating these stages often do not own the factories that manufacture their products. This is the fabless model described by the OECD: design is separated from capital-intensive fabrication and from outsourced assembly and testing (OECD, 2025).

**Architecture, intellectual property, and design software.** Modern chips contain billions of transistors and cannot be designed without electronic design automation (EDA) software used for circuit design, simulation, verification, and preparation of the physical layout. Synopsys, Cadence Design Systems, and Siemens EDA are the principal global commercial suppliers. Reusable processor cores, interfaces, and other IP blocks reduce development time; Arm is the most important commercial licensor of CPU architectures and cores, while open RISC-V specifications support a growing alternative ecosystem. This stage requires less physical capital than fabrication but captures a large share of value added and is difficult to replace because tool chains, libraries, engineering skills, and fabrication-process specifications must be qualified together (OECD, 2025).

**AI-processor design.** NVIDIA is the central merchant supplier of data-centre GPUs and complete accelerated-computing platforms. AMD is its principal merchant competitor in advanced AI accelerators, while Intel supplies CPUs and AI accelerators and retains integrated manufacturing capabilities. Broadcom and Marvell are important in custom application-specific integrated circuits and connectivity. At the same time, the largest cloud companies increasingly act as 'system firms' and design chips for their own services: Google develops TPUs, Amazon Web Services develops Trainium and Inferentia, Microsoft develops Maia, and Alibaba and Huawei develop accelerators for their ecosystems. This vertical integration can reduce dependence on a merchant supplier, but it also strengthens companies that already control cloud infrastructure, software, customers, and distribution (European Commission, 2024; NVIDIA, 2026).

**Manufacturing equipment.** A completed design can be manufactured only with highly specialised machinery. ASML supplies the extreme-ultraviolet lithography systems required for the most advanced process nodes and is therefore a particularly clear chokepoint. Applied Materials, Lam Research, Tokyo Electron, and KLA are leading suppliers of deposition, etching, cleaning, thermal-processing, inspection, metrology, and process-control equipment. These firms are complementary rather than interchangeable: a fabrication plant requires complete, qualified production lines, service engineers, spare parts, and continual process adjustment. Access to one class of machine is not sufficient if another essential class is unavailable (ASML, 2026; OECD, 2025).

**Materials and consumables.** Fabrication also relies on ultra-pure silicon wafers, photoresists, masks, gases, chemicals, and polishing materials. Shin-Etsu Chemical and SUMCO are major suppliers of silicon wafers; Japanese firms including JSR, Tokyo Ohka Kogyo, and Shin-Etsu are important in photoresists and related materials; and Linde, Air Liquide, and Air Products supply high-purity industrial gases. The strategic significance of these companies is not always visible in final-product statistics. A low-cost chemical may nonetheless be a binding constraint if it must meet exceptionally demanding purity standards and requires lengthy customer qualification (OECD, 2023b; OECD, 2025).

**Wafer fabrication.** Taiwan Semiconductor Manufacturing Company (TSMC) is the leading dedicated foundry and the principal manufacturer of many advanced processors designed by fabless firms. Samsung Electronics is the other major company combining leading-edge logic fabrication with large-scale memory production, while Intel is attempting to expand its external foundry business alongside production for its own products. GlobalFoundries, United Microelectronics Corporation (UMC), and Semiconductor Manufacturing International Corporation (SMIC) are large producers at mature or selected advanced nodes. Mature-node capacity remains essential for power management, networking, vehicles, industrial equipment, and the many supporting chips surrounding an AI accelerator. TSMC reported that technologies at 7 nanometres and below generated 74 per cent of its wafer revenue in 2025, illustrating how strongly advanced demand is concentrated in its production system (TSMC, 2026).

**High-bandwidth memory.** Training and serving large models require processors to exchange data rapidly with memory. High-bandwidth memory (HBM) stacks several DRAM dies and places them close to the accelerator, making memory capacity, yield, and packaging as important as the compute die itself. SK hynix, Samsung Electronics, and Micron are the three major suppliers. NVIDIA's own 2026 filing identifies these firms as its memory suppliers, demonstrating that even the leading accelerator designer depends on a small group of external manufacturers (NVIDIA, 2026).

**Advanced packaging, assembly, and testing.** AI systems increasingly combine compute dies, memory stacks, interposers, and high-speed interfaces in one package. TSMC's CoWoS technology is particularly important for advanced accelerators; TSMC also develops SoIC and other three-dimensional integration technologies. ASE Technology, Amkor Technology, Samsung, and JCET are major providers of assembly and testing services, while Japanese suppliers such as Ibiden and Shinko Electric are important in package substrates. Packaging is no longer a low-technology final step: it determines bandwidth, power use, thermal performance, and the usable yield of a system containing several expensive dies (NVIDIA, 2026; TSMC, 2026).

**Boards, servers, and networking.** Packaged processors are mounted on accelerator boards and integrated into servers and racks. Hon Hai Precision Industry (Foxconn), Quanta Computer, Wiwynn, and Wistron are major Taiwanese original-design and contract manufacturers; Dell Technologies, Hewlett Packard Enterprise, and Supermicro are prominent branded system vendors. Large AI clusters also require switches, network adapters, optical connections, and interconnect software. NVIDIA, through its Mellanox-derived networking business, competes and cooperates with Broadcom, Arista Networks, Cisco, AMD, Intel, and Marvell. A shortage of switches, optical modules, power equipment, or cooling can therefore delay usable computing capacity even when GPUs have already been delivered (NVIDIA, 2026).

**Cloud and data-centre deployment.** The final stage converts hardware into computing services. Amazon Web Services, Microsoft Azure, and Google Cloud are the largest global hyperscale platforms; Oracle Cloud is an important supplier of large AI clusters, while Alibaba Cloud, Huawei Cloud, and Tencent Cloud have substantial positions in China and parts of Asia. These companies purchase accelerators and also design their own chips, operate proprietary software stacks, and allocate scarce computing capacity among internal services and external customers. Control over cloud access can therefore become a bottleneck independent of physical chip ownership (European Commission, 2024; NVIDIA, 2026).

A stylised NVIDIA accelerator illustrates the degree of interdependence. The architecture and software ecosystem are controlled by NVIDIA, the physical design is prepared with specialised EDA tools and third-party IP, the compute die may be fabricated by TSMC, HBM is supplied by SK hynix, Samsung, or Micron, the components are joined through CoWoS packaging, and final systems are assembled by contract manufacturers such as Hon Hai or Wistron. The resulting rack is then connected with networking products and installed in a hyperscaler's data centre. NVIDIA's annual report explicitly identifies TSMC and Samsung as foundries, the three major memory manufacturers as suppliers, CoWoS as a packaging technology, and Hon Hai, Wistron, and Fabrinet as contractors (NVIDIA, 2026). The commercial product is thus the outcome of coordinated capacity reservations across several companies and jurisdictions, not the output of one vertically integrated factory.

The geography of the chain is correspondingly specialised. The United States is especially strong in accelerator design, EDA, semiconductor equipment, cloud services, and software. Taiwan is central to advanced foundry production, packaging, and server manufacturing. South Korea is central to memory and also operates advanced logic fabs. Japan is a leading source of wafers, chemicals, photoresists, substrates, and manufacturing equipment. The Netherlands hosts ASML. China has large capacity in mature-node fabrication, assembly and testing, electronics manufacturing, cloud services, and a rapidly developing domestic equipment and design ecosystem, but it remains constrained in several advanced segments. The OECD estimates that five economies account for about three-quarters of global semiconductor value added, while also warning that aggregate data can conceal still higher concentration in specific non-substitutable products (OECD, 2023b; OECD, 2025).

This structure creates several distinct forms of power. A government can restrict exports of advanced processors, manufacturing equipment, EDA tools, or specialised materials; control access to cloud computing; screen investment; or apply sanctions to firms and their customers. A company can exercise power through capacity allocation, licensing terms, technical interfaces, proprietary software, or the decision to support particular customers and jurisdictions. Standards and developer ecosystems create an additional layer: replacing a chip may require rewriting software and retraining engineers even where a nominal hardware substitute exists. Chokepoint power therefore depends not only on market share, but also on switching costs, qualification periods, inventories, and the time required to construct new capacity.

The term 'supremacy', however, should be avoided or carefully defined. Control over one bottleneck does not eliminate dependence elsewhere. The United States hosts many leading design, software, equipment, and cloud companies but relies heavily on Asian fabrication, memory, packaging, and electronics manufacturing. Taiwan's leading foundries depend on equipment, software, materials, energy, water, and foreign customers. South Korean memory producers depend on imported machinery and chemicals. ASML itself depends on a network of specialised optical, laser, mechanical, and electronic suppliers. The result is asymmetric interdependence, not complete autonomy.

Chokepoint power is also dynamic. Export restrictions or supply shocks can impose immediate costs because alternative suppliers and process routes cannot be qualified quickly. Over a longer horizon, however, they encourage stockpiling, substitution, domestic research, industrial subsidies, and the construction of parallel ecosystems. Governments in the United States, the European Union, China, Japan, South Korea, and Taiwan have consequently supported new fabrication and packaging capacity. These policies may improve resilience, but they can also produce subsidy competition, excess capacity in some segments, higher production costs, and fragmentation of technical standards and markets.

Resilience should therefore not be equated with complete national self-sufficiency. A more realistic strategy combines supplier diversification, geographic dispersion, transparent capacity and dependency mapping, inventories of genuinely non-substitutable inputs, reciprocal agreements with trusted partners, workforce development, and the ability to move workloads between providers. Diversification must be assessed stage by stage: adding a second chip designer does little if both depend on the same foundry, packaging line, memory supplier, cloud platform, or energy-constrained region.

A further division is emerging between countries capable of developing or hosting advanced AI infrastructure and countries that primarily import models and cloud services. Dependence on foreign providers may constrain policy autonomy even without formal political coercion. Digital sovereignty should therefore be understood not as autarky, but as the practical ability to change suppliers, protect data, audit systems, maintain essential services during disruption, and negotiate access on acceptable terms. In this sense, power in the AI economy belongs not only to the companies that design the most visible models or processors, but also to the less visible firms that control the tools, materials, memory, packaging, networks, and infrastructure on which those systems depend.

## 6. AI-Enabled Robotics and Traditional Industrial Employment

The effects of AI-enabled robotics should be distinguished from those of generative AI. Generative systems primarily transform cognitive and communicative tasks, whereas robotics connects computation to perception and physical action. The relevant system is therefore not simply a model, but a model combined with cameras, microphones, force and position sensors, actuators, connectivity, and an organisation that determines the machine's objectives. This embodiment creates benefits that software alone cannot deliver, but it also converts errors, cyberattacks, and managerial choices into events that may affect bodies, workplaces, homes, and public space.

Mass adoption is no longer a purely speculative scenario, although today's evidence still concerns mostly conventional industrial and logistics robots rather than general-purpose humanoids. The International Federation of Robotics reported 4.664 million industrial robots in operational use in 2024, 9 per cent more than a year earlier, and 542,000 new installations during that year. China accounted for 54 per cent of new deployments, and the IFR expects annual global installations to exceed 700,000 by 2028 (IFR, 2025). These figures establish the scale of factory automation, but they should not be read as a count of fully autonomous AI systems.

Traditional industrial robots were best suited to repetitive operations in structured environments. AI expands the range of feasible activities by enabling machine vision, adaptive control, autonomous navigation, predictive maintenance, natural-language instruction, and learning from demonstration. Automation can consequently spread from welding, painting, and machine loading to warehouses, food processing, agriculture, construction, cleaning, inspection, delivery, health care, and domestic assistance. The important social change is not only that more tasks become technically automatable, but also that robots can operate closer to people and in environments that are less predictable than a fenced production cell.

The central analytical distinction remains the difference between automating tasks and eliminating occupations. Most jobs combine physical execution with setup, maintenance, quality control, exception handling, communication, safety, and judgement. A robot can remove a bundle of routine activities while leaving the occupation intact, redesign the occupation around supervisory work, or make the remaining position more demanding. Technical capability is therefore only one determinant of adoption; relative cost, product demand, liability, regulation, worker acceptance, and the cost of reorganising the workplace are equally important.

The risk of direct displacement is greatest where work is repetitive, predictable, measurable, physically demanding, performed at high volume, and embedded in a controlled environment. It is lower where tasks require irregular movement, tacit knowledge, responsibility for ambiguous outcomes, social trust, or frequent adaptation to exceptional circumstances. AI shifts this boundary, but it does not abolish it. Even an apparently autonomous system depends on installation, remote supervision, cleaning, repair, data annotation, software updates, energy, and human intervention when its confidence or physical capability is insufficient.

Evidence from the United States shows that displacement can be substantial in exposed local labour markets. Acemoglu and Restrepo estimate that one additional industrial robot per thousand workers between 1990 and 2007 reduced the employment-to-population ratio by 0.18-0.34 percentage points and wages by 0.25-0.5 per cent (Acemoglu and Restrepo, 2020). These coefficients describe a particular historical setting and should not be exported mechanically to other countries or to future humanoid robots. They nevertheless show that productivity growth at adopting firms can coexist with measurable losses for workers in the surrounding labour market.

Germany provides a different adjustment pattern. Dauth et al. estimate that each additional robot displaced roughly two manufacturing jobs, but total local employment did not fall because service employment expanded. Much of the adjustment occurred through fewer young workers entering manufacturing rather than immediate dismissal of incumbent employees. Workers who remained with their original employers were partly protected, although some experienced lower cumulative earnings (Dauth et al., 2021). The case demonstrates that employment stability can conceal occupational closure for a new generation and a gradual transfer of work to sectors with different wages and institutions.

Japanese evidence further cautions against a universal job-loss coefficient. Adachi, Kawaguchi, and Saito exploit changes in robot prices and find that cheaper robots increased both robot use and employment by raising productivity and the scale of output in adopting industries (Adachi et al., 2024). Whether the productivity effect offsets displacement thus depends on demand, export performance, domestic supply chains, and the speed with which production expands. A robot may substitute for labour per unit of output while complementing labour at the level of an expanding firm or industry.

Three mechanisms therefore determine the aggregate outcome. The displacement effect reduces labour required for tasks now performed by machines. The productivity and scale effect lowers costs, improves quality, and can raise demand enough to preserve or expand employment. The new-task effect creates work in integration, maintenance, safety, process design, data analysis, customer interaction, and handling exceptional cases (Acemoglu and Restrepo, 2018; Restrepo, 2023). The balance is not technically predetermined; it is shaped by competition, demand, skills, labour-market institutions, and the extent to which cost savings are translated into lower prices, higher output, wages, or profits.

Even a neutral national employment balance may therefore produce serious social disruption. New jobs may emerge in different firms, sectors, regions, and years and may require qualifications that displaced workers do not possess. Experienced production workers often hold valuable but firm- or sector-specific knowledge. If a plant automates rapidly, the relevant comparison is not between the number of jobs destroyed and created in the entire economy, but between the worker's former wage and security and the jobs realistically accessible without relocation, a long period of training, or a large income loss.

Distributional effects can be cumulative. Owners of robotic capital, successful technology suppliers, and workers whose skills complement automation may receive higher income, while workers performing substitutable tasks face weaker bargaining power. If displaced workers move into lower-paid services, labour supply in those sectors can also increase and place pressure on wages there. Models of automation consequently predict that productivity growth can coexist with rising income and wealth inequality when ownership is concentrated and new human tasks are not created rapidly enough (Moll et al., 2022). Differences by age, education, gender, disability, and migration status depend on the task structure of particular industries; they should be measured rather than assumed.

Regional concentration magnifies these effects. A town organised around one automotive, metal, or logistics employer cannot diversify as easily as a national economy. Bekhtiar's study of Austria finds that robotisation reduced manufacturing employment and increased out-migration from exposed rural areas; it accounted for roughly one-quarter of rural-to-urban net migration between 2003 and 2016, driven mainly by young and medium- or low-skilled people (Bekhtiar, 2025). The resulting loss of population and tax revenue can weaken schools, transport, retail, housing demand, and local public services, creating a feedback loop in which the most mobile residents leave and the adjustment burden falls on those who remain.

The social effect of robotics also depends on the quality of the jobs that survive. AI-supported robots can remove heavy lifting, toxic exposure, monotonous inspection, and dangerous machine contact. They can equally be used to decompose work into tightly measured actions, set machine-determined production rhythms, and monitor workers continuously. In OECD surveys, one in five AI users reported reduced control over the sequence of their tasks; workers exposed to algorithmic management were more likely to report increased work intensity, and most workers whose employers collected data about them expressed privacy concerns (OECD, 2023a). Research across 20 European countries similarly associates robotisation with lower perceived work meaningfulness and autonomy (Nikolova et al., 2024).

Safety effects are therefore mixed rather than uniformly positive. Using United States establishment data, Gihleb et al. find that a one-standard-deviation increase in robot exposure reduced work-related injuries by about 16 per cent, with larger reductions in manufacturing; German data also show lower physical intensity and disability (Gihleb et al., 2022). Yet a study of warehouse robotics finds a 40 per cent reduction in severe injuries alongside a 77 per cent increase in non-severe injuries, partly because the pace of the remaining human work accelerated (Burtch et al., 2025). Automation can remove the most dangerous task while intensifying the tasks left to people.

Job insecurity and loss of control can also affect mental health before displacement occurs. Abeliansky, Beulmann, and Prettner find that greater exposure to industrial robots worsened the mental health of German workers, with fear of job loss and labour-market polarisation acting as transmission channels (Abeliansky et al., 2024). This does not imply that proximity to a robot is intrinsically harmful. It indicates that uncertainty about employability, earnings, and status is itself a social cost and that the manner in which adoption is announced and negotiated matters.

The most severe evidence concerns community-level health in the United States. O'Brien, Bair, and Venkataramani estimate that each additional robot per thousand workers was associated with just over eight additional deaths per 100,000 men aged 45-54 and just under four additional deaths per 100,000 women of the same age. The study links automation exposure to drug-overdose and suicide mortality and finds larger effects in manufacturing-intensive areas; the authors also report suggestive moderation by more generous unemployment insurance (O'Brien et al., 2022). These are historical causal estimates for exposed US communities, not a forecast that every new robot will produce the same health effect. Their importance lies in showing that labour-market shocks can become public-health shocks when income, identity, and local institutions deteriorate together.

Household and demographic effects may follow the same pathway. In US commuting zones, Anelli, Giuntella, and Stella associate robot exposure with lower marriage and marital fertility, and with higher divorce, cohabitation, and non-marital births. The proposed mechanism is not a cultural response to machines themselves, but a change in the relative employment and earnings prospects of men and women combined with greater uncertainty (Anelli et al., 2024). Such findings should be treated as context-dependent, but they broaden the relevant unit of analysis from the displaced employee to the household.

Economic insecurity can also become political discontent. Across 13 Western European countries, individual vulnerability to industrial robot adoption was associated with greater support for radical-right parties (Anelli, Colantone, and Stanig, 2021). The result does not mean that robotics mechanically produces extremism. It suggests a pathway in which concentrated losses, perceived status decline, and distrust of institutions generate political polarisation when adjustment policy appears ineffective or unfair. Regions that gain productive investment and regions that lose accessible employment may consequently develop sharply different attitudes toward technology, trade, and democratic institutions.

Robots used in care create a different balance of benefits and risks. They can support mobility, medication reminders, rehabilitation, monitoring, communication with relatives, and repetitive physical assistance, while easing labour shortages and reducing strain on caregivers. A 2025 meta-analysis covering 19 studies and 1,083 older adults found that social robots reduced loneliness, with stronger effects in institutional settings (Mehrabi and Ghezelbash, 2025). The evidence supports robots as a possible supplement to care, not as proof that automated companionship is an adequate substitute for sustained human relationships.

If providers use robots primarily to reduce staffing costs, assistance may become technically continuous but socially thinner. Ethical concerns include loss of human contact, infantilisation, deception about the machine's emotional capacities, reduced freedom to refuse monitoring, and ambiguous consent among people with cognitive impairment (Sharkey and Sharkey, 2012). A companion robot can reduce loneliness when it adds interaction, yet increase institutional neglect if its presence is used to justify fewer visits by carers or relatives. The relevant counterfactual is therefore decisive: replacing isolation with a robot may be beneficial, whereas replacing good human care with a robot may not be.

Privacy, cybersecurity, and responsibility become more important as robots enter homes and public services. Mobile systems may continuously collect images, voices, movements, health information, and spatial maps. Because a robot also has actuators, unauthorised access or a mistaken perception can produce physical as well as informational harm. Reviews of robotic cybersecurity identify risks involving authentication, communication links, software updates, sensors, cloud services, and malicious control (Yaacoub et al., 2022). Mass deployment also complicates liability: responsibility may be divided among the model developer, robot manufacturer, component supplier, system integrator, owner, employer, cloud provider, and human supervisor.

Access to beneficial robotics may itself become a source of inequality. Large firms can finance integration and reorganise processes, while smaller firms may remain less productive or become dependent on robotics-as-a-service providers. Wealthier households and well-funded care institutions may purchase assistive systems that extend independence, while poorer users receive lower-quality, more intrusive, or mandatory automated services. People who cannot use speech interfaces, whose language or accent is poorly represented, or whose homes are unsuitable for the device may be excluded. The distributional question is therefore not only who loses work, but also who receives safer jobs, better care, and greater autonomy.

**A scenario-based estimate of scale.** No credible study can yet state how many social problems general-purpose AI robots will create. The technology, price trajectory, regulation, and organisational response are too uncertain, and existing evidence largely concerns industrial robots. A defensible estimate should therefore combine a transparent adoption scenario with empirically observed ranges and a qualitative assessment of risks for which no population-level data exist.

For industrial adoption, a simple reference scenario starts from the 4.664 million robots operating in 2024. If the operational stock grew by 7-9 per cent annually, it would reach approximately 7.0-7.8 million units in 2030. This is an arithmetic scenario rather than an IFR forecast: it ignores retirements, recessions, capacity constraints, and the difference between conventional and AI-enabled robots. It is nevertheless consistent with the IFR's expectation that annual installations will exceed 700,000 by 2028 and indicates that exposure will broaden even without a breakthrough in humanoid robotics (IFR, 2025).

For employment and earnings, a rapid but still historically recognisable local shock can be represented by an increase of two robots per thousand workers. Applying the US estimates mechanically gives a benchmark of a 0.36-0.68 percentage-point reduction in the employment-to-population ratio and a 0.5-1.0 per cent wage reduction in exposed local labour markets (derived from Acemoglu and Restrepo, 2020). These values are not global forecasts. Germany and Japan show that sectoral reallocation or output expansion can offset aggregate losses. They are best interpreted as a plausible downside range for places with weak demand growth, limited worker mobility, and inadequate transition policy.

A broader task-based OECD estimate provides an outer envelope for all forms of automation, not robotics alone: about 14 per cent of jobs were judged highly automatable and another 32 per cent likely to change substantially over 10-20 years (OECD, 2019b). The socially relevant population is thus likely to be much larger than the number of workers who become unemployed. Many more people may face retraining, closer monitoring, loss of occupational identity, altered career ladders, or pressure on wages even while remaining employed.

For health and community effects, the US historical evidence implies that the same two-robot exposure benchmark was associated with roughly 16 additional deaths per 100,000 men aged 45-54 and somewhat fewer than eight among women of the same age (O'Brien et al., 2022). This estimate must not be universalised: German results are less adverse, and social insurance, health care, labour relations, and local economic diversification differ substantially. It should be treated as a high-severity warning for communities where automation arrives as an unmanaged employment shock.

For care, domestic service, autonomous mobility, and public-space robotics, numerical projections are much weaker. The installed base is smaller, systems are heterogeneous, and trials are often short. The most plausible near-term outcome is not wholesale replacement of carers or service workers, but partial automation combined with remote supervision and data-intensive management. The principal measurable risks will therefore initially be changes in staffing ratios, frequency of human contact, consent and privacy incidents, service exclusion, accident rates, and the distribution of access, rather than a single employment total.

**High probability and broad reach: task reorganisation.** If adoption continues, millions of workers are likely to experience changed task content, training requirements, performance measurement, or career entry routes. This is more probable than economy-wide technological unemployment and may affect a substantial share of occupations even where headcount remains stable.

**High probability but concentrated harm: displacement and wage loss.** The largest losses are likely to cluster in routine-intensive firms and manufacturing or logistics regions. National averages will understate the severity for particular workers, age cohorts, subcontractors, and towns.

**Medium-to-high probability: inequality and weaker bargaining power.** Without broader ownership, competition, collective bargaining, or redistribution, the financial gains are likely to accrue disproportionately to capital owners, technology suppliers, complementary professionals, and large firms, while transition costs are borne locally.

**Medium probability but potentially high severity: health, family, and political spillovers.** These effects are indirect and institution-dependent, but evidence on mortality, household formation, migration, and voting shows that they cannot be dismissed as speculative externalities of the labour market.

**Mixed net effect: occupational safety and care quality.** Robots can remove hazardous work and reduce loneliness, but poorly designed systems can accelerate work, create new injuries, displace human contact, and normalise intrusive monitoring. Outcomes depend strongly on workflow and staffing decisions.

**Lower-frequency but high-consequence risk: physical failure and cyberattack.** Population-level probabilities are not yet measurable, but mass deployment increases the number of connected physical systems whose failures can affect people directly. Safety certification, secure updates, incident reporting, and clear liability become essential public institutions rather than optional product features.

The most plausible social problem is therefore not a sudden world without work. It is an uneven transition in which productivity rises while particular workers lose earnings and status, entry-level pathways narrow, some regions decline, surveillance and work intensity increase, and essential care or public services become dependent on systems that users cannot meaningfully inspect or refuse. At the same time, robotics can reduce injuries, compensate for labour shortages, extend independent living, and create new forms of skilled work. The balance between these outcomes depends on deployment choices and on the institutions discussed in the next chapter.

## 7. Social Protection and the Role of Trade Unions

It would be inaccurate to claim that trade unions universally fail to recognise the risks of AI. The more defensible conclusion is that their responses are uneven and still cover only a minority of workers. A European study of services-sector unions found that 20 per cent reported an organisation- or sector-level collective agreement addressing AI, while 42 per cent were engaged in related social dialogue or negotiations (Eurofound, 2025). Falling union membership, fragmented platform work, limited technical expertise, and informational asymmetry between employers and workers all restrict the reach of collective action. Where institutions are strong, however, recent cases show that AI deployment can become a negotiable organisational choice rather than a unilateral technical decision.

Trade unions and works councils matter because many consequences described in Chapter 6 are determined below the level of national legislation. The same system can eliminate positions, assist workers, intensify work, or create new career paths depending on staffing ratios, performance targets, data collection, training, and the distribution of productivity gains. Regulation can establish minimum rights, but collective bargaining can translate them into enforceable rules for a particular occupation, production process, or workplace.

**Collective bargaining backed by industrial action.** The 2023 Writers Guild of America strike made generative AI a central bargaining issue rather than treating it as a matter of managerial discretion. The resulting Minimum Basic Agreement states that AI-generated text is not literary or source material, so it cannot be used to reduce a writer's credit or compensation. A writer may use AI with company consent but cannot be required to do so, and a company must disclose when material supplied to a writer was generated by AI. The agreement leaves unresolved whether studios may use writers' works to train models, but it demonstrates that collective bargaining can protect occupational authorship and bargaining-unit work without prohibiting voluntary use of the technology (WGA, 2023).

SAG-AFTRA used a 2023 strike to obtain a different set of safeguards suited to performers. Its television and theatrical agreement requires informed consent for creating and using a performer's digital replica, generally requires compensation and residuals, and gives the union notice and an opportunity to bargain over synthetic performers. Producers may not use replicas simply to avoid hiring background performers or evade contractual coverage limits (SAG-AFTRA, 2023). These clauses convert likeness and voice from data that might be captured once and reused indefinitely into continuing subjects of consent, payment, and representation. Later agreements extended similar protections to animation, sound recordings, and video games.

The approach has diffused beyond Hollywood. By May 2025, the NewsGuild-CWA reported that more than three dozen newsroom agreements contained AI language. Stronger clauses protected bargaining-unit work, required journalistic oversight of AI-assisted output, and established enforceable limits on employer-directed use. Members used internal organising, public campaigns such as Politico's 'Journalists, Not Robots', and in some cases strike action to obtain these terms (NewsGuild-CWA, 2025). The example also reveals a limitation: contractual protections are strongest in organised newsrooms, while freelancers, subcontractors, and non-union employees may remain outside their scope.

A particularly relevant platform-work example is the 2024 Hilfr2 agreement between the Danish cleaning platform Hilfr and the trade union 3F. The agreement classifies cleaners as employees, makes the company responsible for algorithmic decisions, and allows such decisions to be challenged through the Danish labour-dispute system. It also places a link on the platform to a union 'digital clubhouse' where dispersed workers can obtain advice, elect representatives, and discuss conditions without employer oversight (Eurofound, 2025). This addresses a central organisational problem of digital labour: workers managed through the same platform may rarely meet one another physically.

Other European agreements show that bargaining can operate above company level. A 2024 declaration between the European Banking Federation and UNI Europa addressed responsible AI in banking. Italy's 2024 cross-industry agreement added AI-related occupations and stressed impact assessment, information, education, and retraining. Belgium's banking and insurance discussions linked AI training to sectoral training funds and to an older collective agreement requiring written information and consultation when new technology has significant effects on employment or working conditions. Spain's national cross-industry agreement called for understandable information about AI used in recruitment, evaluation, promotion, and dismissal (Eurofound, 2025). These arrangements spread the cost of expertise and training across a sector, which can be especially important for smaller employers.

**From bargaining demands to statutory rights.** Spain's Riders Law illustrates how trade-union pressure and social dialogue can be translated into generally applicable law. Ley 12/2021 created a presumption of employment for many delivery-platform workers and amended the Workers' Statute so that employee representatives must be informed about the parameters, rules, and instructions underlying algorithms or AI systems that influence working conditions, access to employment, and job retention, including profiling (Spain, 2021). The law does not require disclosure of source code and does not by itself make a complex system understandable. Its importance lies in recognising algorithmic management as a collective labour issue rather than solely an individual data-protection question.

The EU Platform Work Directive extends several of these principles across the Union. It restricts certain forms of data processing, requires information about automated monitoring and decision-making, provides for human oversight and review of significant decisions, and gives representatives consultation and information rights. Member States must transpose the directive into national law by December 2026 (European Parliament and Council, 2024). Its coverage is narrower than a general right governing all workplace AI, but it establishes a regulatory floor below which platform companies cannot contract and gives unions tools that do not depend entirely on achieving recognition at each firm.

**Joint company-union programmes.** The Microsoft-AFL-CIO partnership announced in 2023 offers a cooperative model alongside adversarial bargaining. It has three stated objectives: educating workers and union leaders about AI, incorporating worker expertise into the development of AI systems, and jointly shaping public policy concerning frontline workers' skills. It also builds on Microsoft's labour-neutrality commitments for organising in parts of its business (Microsoft and AFL-CIO, 2023). Such a partnership can give unions earlier access to technical knowledge, but its value must ultimately be assessed through concrete changes in product design, workplace deployment, training access, and negotiated employment outcomes rather than by the existence of dialogue alone.

**Company-led redeployment.** IKEA provides a useful example of task substitution combined with internal reassignment. After the Billie chatbot began handling routine customer queries, Ingka Group reported that it had trained 8,500 call-centre employees as remote interior-design advisers between 2021 and 2023. Billie handled 47 per cent of call-centre queries over the same period, while remote interior-design sales generated EUR 1.3 billion in the 2022 financial year (Reid, 2023). This is evidence that automation can be linked to a higher-value service and a new revenue stream rather than immediate redundancy. It is not proof of a universal solution: the new work existed because the company had a complementary service to expand, employees could acquire the relevant skills, and customer demand was sufficient.

**Tripartite transition policy.** Singapore's Company Training Committee model links public finance, employer investment, and worker representation. Companies applying for the NTUC CTC Grant form a committee with management and worker representatives, develop a transformation plan, and commit to worker outcomes such as wage increases, career-development plans, or skills allowances. The grant has co-financed up to 70 per cent of eligible transformation and associated training costs. By September 2025, more than 700 projects at over 500 companies had been approved; more than 70 involved AI, and almost 10,000 workers were expected to benefit through an average wage increase of 5 per cent above their annual increment, structured career paths, or skills allowances (NTUC, 2025). The design is notable because subsidy eligibility is tied to verifiable worker outcomes rather than technology purchase alone.

The SBS Transit case shows how this model works at workplace level. The company and the National Transport Workers' Union used their training committee to prepare employees for digitalisation, electric vehicles, and autonomous buses. An AI-supported tyre-management system was designed to automate inspection and predictive maintenance, freeing the equivalent of one technician per workshop for other work. The accompanying programme created a Diagnostic Expert career scheme, a skills roadmap, and enhanced wage prospects for more than 50 workers; bus captains moving into autonomous-vehicle-related roles were also promised improved progression and remuneration (NTUC, 2025). The example connects deployment, job redesign, certification, and pay instead of treating training as a stand-alone course.

**Public training and advisory infrastructure.** Germany's response to digital and ecological transformation combines the National Skills Strategy with regional and workplace-level support. The federal labour ministry's Zukunftszentren, or transformation hubs, provide tailored advice and training, while the New Quality of Work Initiative supports companies and employees in testing ways to preserve good working conditions. The programmes were developed with the Federal Employment Agency, the Länder, employers, trade unions, and other social partners (BMAS, 2023). This institutional infrastructure is less visible than a single corporate initiative, but it can reach small and medium-sized firms that lack their own training departments and can connect technological change to regional labour-market policy.

**Why retraining is not enough.** These cases should not be interpreted as proof that every displaced worker can be converted into an AI specialist. Training is effective only when suitable jobs exist, workers can participate during paid time or with income support, programmes correspond to actual labour demand, and employers recognise the resulting qualifications. Selection also matters: firms may offer intensive programmes to workers already most likely to succeed while lower-paid, older, disabled, temporary, or subcontracted workers receive only generic digital-literacy courses. Evaluation should therefore track completion, internal placement, wage progression, job retention, and access by employment status, age, gender, and region, not merely enrolment or training hours.

A credible organisational agreement should operate before, during, and after deployment. Before adoption, employers should notify representatives, identify affected tasks and groups, disclose the purpose and data sources of the system, and conduct employment, equality, privacy, and safety assessments. During implementation, workers and their representatives should be able to test the system, obtain independent technical advice, challenge unrealistic targets, and require human review of disciplinary, scheduling, promotion, or dismissal decisions. After deployment, the parties should monitor errors, workload, injuries, staffing, wage effects, and whether promised redeployment has occurred. These procedures turn consultation from a one-off presentation into continuing governance.

Social protection must also cover cases in which internal redeployment fails. Relevant instruments include unemployment protection, wage insurance for workers who accept lower-paid jobs, mobility and housing assistance, pension-credit protection, employer-financed transition funds, and targeted investment in regions dependent on exposed industries. Shorter working hours with maintained hourly pay can distribute productivity gains when output can be preserved with fewer labour hours; negotiated gain-sharing or profit-sharing can perform a similar distributive function. The appropriate mix depends on the duration of displacement and whether the shock affects one company, an occupation, or an entire local economy.

The objective should not be to prevent all automation. Blocking technologies that raise productivity or remove dangerous work could preserve inefficient and harmful tasks. The practical lesson from the cases above is instead that beneficial adjustment requires institutions capable of bargaining over the design and timing of change, enforcing transparency and human review, financing training and transition, and sharing measurable gains. Where worker representation is absent, voluntary company programmes may still help, but workers depend on managerial commitment and have limited recourse if priorities change. AI policy should therefore strengthen both universal minimum rights and the collective capacity to negotiate above that floor.

## 8. Conclusion

Artificial intelligence should be analysed as a material and institutional transformation rather than as an abstract digital technology. Its development requires capital, electricity, water, semiconductors, land, specialised labour, and international supply networks. The distribution of these resources is already reshaping corporate competition and geopolitical relations.

AI-related investment may increase productivity but also reinforce market concentration. Data-centre expansion may support innovation while creating local electricity and water conflicts. Control over semiconductor and cloud-computing bottlenecks may generate geopolitical leverage while simultaneously exposing controlling states to reciprocal dependencies. AI may contribute to environmental protection, but its benefits do not erase the lifecycle footprint of computing infrastructure.

In the labour market, AI-enabled robotics is likely to reduce demand for some traditional industrial occupations, particularly those dominated by routine physical tasks in structured environments. This does not imply the disappearance of industrial labour or an inevitable decline in total employment. Productivity growth, increased production, and the creation of new tasks may offset part of the displacement. Nevertheless, aggregate compensation does not protect individual workers or regions from lasting losses.

The central policy issue is therefore not whether AI is intrinsically beneficial or harmful. It is whether societies can govern its development so that infrastructure costs, environmental burdens, geopolitical dependencies, and labour-market risks are not imposed on groups with the least capacity to influence technological decisions. Competition policy, environmental accounting, resilient supply chains, social protection, collective bargaining, and worker participation should be regarded as essential components of AI policy rather than as secondary responses to technological change.